\documentclass[11pt]{article}
\usepackage{amsmath,amssymb,color,graphics,epsfig,cite,comment}
\usepackage[font=small]{caption}
\usepackage{amsfonts}
\graphicspath{{fig/}}
\usepackage{hyperref} 
\hypersetup{
colorlinks=true,
linkcolor=blue, % 内部链接颜色
citecolor=blue, % 引用链接颜色
urlcolor=blue, % URL链接颜色
pdfborder={0 0 0} % 去除链接边框（如果colorlinks=false）
}
\newcommand{\be}{\begin{equation}}
\newcommand{\ee}{\end{equation}}
\newcommand{\bea}{\setlength\arraycolsep{2pt} \begin{eqnarray}}
\newcommand{\eea}{\end{eqnarray}}

\def\0{{\sst{(0)}}}
\def\1{{\sst{(1)}}}
\def\2{{\sst{(2)}}}
\def\3{{\sst{(3)}}}
\def\4{{\sst{(4)}}}
\def\5{{\sst{(5)}}}
\def\6{{\sst{(6)}}}
\def\7{{\sst{(7)}}}
\def\8{{\sst{(8)}}}
\def\sst#1{{\scriptscriptstyle #1}}

\begin{document}

\begin{center}
{\Large {\bf Scalarization of Charged Black Hole in Gauss-Bonnet Extended Starobinsky-Maxwell Gravity } }

\vspace{20pt}

{\large Rui-Xi Zhu, Hai-Shan Liu }

\vspace{10pt}

\vspace{10pt}

{\it Center for Joint Quantum Studies and Department of Physics,\\
School of Science, Tianjin University, Tianjin 300350, China }

\vspace{40pt}

\underline{ABSTRACT}

\end{center}

We re‑examine spontaneous scalarization of black holes within Starobinsky gravity supplemented by the Gauss-Bonnet invariant. A novel feature is uncovered: scalarized solutions split into two smooth branches that connect at a common minimal horizon radius, a multi‑branch structure previously unreported for static, spherically symmetric scalarization. Extending the theory with a Maxwell field, we find that this multi‑branch structure persists, while a new additional disconnected branch emerges within certain parameter ranges. We thoroughly analyse the transition from a single smooth branch to two disconnected branches. Lastly, we verify, employing the standard Maxwell thermodynamic relations, that all charged hairy solutions obey the first law exactly.

\vfill{\footnotesize ruixi\_zhu@tju.edu.cn ~~~ hsliu.zju@gmail.com(corresponding author)  }
%\vfill {\footnotesize mrhonglu@gmail.com}

\thispagestyle{empty}
\pagebreak
%\voffset=-40pt
%\setcounter{page}{1}

%\tableofcontents
%\addtocontents{toc}{\protect\setcounter{tocdepth}{2}}

%\newpage

\section{Introduction}

Black hole scalarization has emerged as a vibrant research frontier, as it furnishes a natural mechanism to evade classical no-hair theorems and yields potentially detectable signatures for gravitational-wave astronomy and black hole imaging. This renders scalarization a powerful and promising probe of strong-field gravitational physics beyond General Relativity (GR). Broadly defined, black hole scalarization is the phenomenon whereby black holes acquire nontrivial scalar hair via the coupling between a scalar field and the gravitational or matter sectors of a modified gravity theory. Depending on the underlying theoretical framework, scalarization can be triggered either by a linear perturbative instability of the scalar-free GR background or by purely nonlinear dynamical effects, giving rise to distinct families of scalarized black hole solutions. Originally discovered in scalar-tensor theories with curvature-dependent couplings, the scalarization mechanism has since been generalized to a wide spectrum of modified gravity models, including Einstein-scalar-Gauss-Bonnet gravity\cite{Antoniou:2017acq,Silva:2017uqg,Doneva:2017bvd}, Einstein-Maxwell-scalar theory\cite{Herdeiro:2018wub,Fernandes:2019rez}, and other nonminimally coupled gravitational frameworks \cite{Liu:2024bzh,Li:2025dpo,Li:2025vfk,Li:2025vcq,Zhang:2026abo}. Scalarized black hole solutions generally bifurcate from the GR black hole solution branch or appear as disconnected solution families, distinguished by nonvanishing scalar charges, altered horizon geometric and thermodynamic properties, and unique quasinormal mode spectra. Notably, most such solutions exhibit dynamical stability across extensive regions of the parameter space.

All previously known realizations of black hole scalarization relied on the presence of an explicit scalar field in the fundamental action. Recently, however, it was shown that scalarization can also emerge in pure higher-curvature gravity, where the scalar degree of freedom responsible for scalarization arises dynamically from the gravitational sector rather than being introduced as an independent matter field. Such a mechanism requires extending Einstein gravity by incorporating higher-curvature corrections. As a natural generalization of General Relativity, higher-curvature gravity is constructed by supplementing the Einstein-Hilbert action with higher-order polynomial invariants of the Riemann tensor, the quadratic extension being the simplest example. In four dimensions, quadratic gravity is perturbatively renormalizable, albeit at the cost of introducing a ghostlike massive spin-2 mode into the spectrum \cite{Stelle:1976gc}. The origin of this ghost is that the field equations generally contain fourth-order derivatives, so that, besides the usual massless graviton, the linearized spectrum also includes a massive scalar mode and a massive spin-2 mode. Various approaches have been proposed to eliminate or circumvent the ghost problem. One possibility is to tune the coupling constants to critical values such that the massive spin-2 mode becomes massless \cite{Li:2008dq,Bergshoeff:2009hq,Liu:2009bk,Deser:2011xc}, although this generally leads to logarithmic ghost modes \cite{Lu:2011zk,Bergshoeff:2011ri,Porrati:2011ku}. Another approach is to consider special combinations of curvature invariants whose field equations remain second order, such as the Gauss-Bonnet term and, more generally, the Lovelock series, which become dynamically nontrivial only in spacetime dimensions $D \geq 5$. In four dimensions, the absence of additional massive excitations in the linearized spectrum around maximally symmetric vacua can instead be achieved in certain classes of massless higher-curvature gravities, including quasitopological gravities and their generalizations \cite{Myers:2010ru,Oliva:2010eb,Bueno:2016xff,Li:2017ncu}.

In generic higher-derivative gravities, the massive spin-2 mode is ghostlike, whereas the massive scalar mode can remain unitary. Consequently, a ghost-free theory can be obtained by imposing appropriate constraints on the coupling constants to decouple the massive spin-2 mode while retaining the scalar degree of freedom. The simplest and most celebrated example is the Starobinsky $R+\alpha R^2$ model \cite{Starobinsky:1980te}. Although Schwarzschild remains an exact solution in this theory, no static black hole can support scalar hair \cite{Nelson:2010ig,Lu:2015cqa}. In contrast, quadratically extended gravity admits new black-hole solutions carrying massive spin-2 hair \cite{Lu:2015cqa}. Along these lines, Liu \textit{et al.} proposed a ghost-free higher-derivative pure-gravity theory, known as Gauss-Bonnet extended Starobinsky gravity, in which the scalar degree of freedom inherited from Starobinsky gravity is nonminimally coupled to the Gauss-Bonnet invariant. They numerically demonstrated that this theory admits scalarized black-hole solutions, providing the first explicit realization of black hole scalarization in a pure higher-curvature gravitational theory \cite{Liu:2020yqa}.

In this work, we first revisit Gauss--Bonnet extended Starobinsky gravity and uncover several new features that have not been reported previously. We then generalize this pure higher-curvature gravity by introducing a Maxwell field and investigate the resulting charged scalarized black holes. We construct the corresponding numerical solutions and study their bifurcation structure, domain of existence, and thermodynamic properties. A detailed comparison between the neutral and charged cases reveals several novel characteristics of scalarized black holes in this framework.

The remainder of this paper is organized as follows. In Sec.2, we briefly review Gauss--Bonnet extended Starobinsky gravity. In Sec.3, we describe the numerical method used to construct the scalarized black hole solutions. In Sec.4, we revisit black hole scalarization in the neutral theory and identify a previously unknown branch of scalarized black hole solutions. In Sec.5, we extend the theory by introducing a Maxwell field and construct novel charged scalarized black holes. Finally, we summarize our results and discuss future directions in Sec.6.

\section{Theory}

A common approach to extending the gravitational theory is to incorporate quadratic curvature invariants into Einstein's theory. Among these extensions, the simplest ghost-free model is provided by adding the quadratic Ricci scalar invariant $R^2$ to the Einstein-Hilbert action, which leads to the Starobinsky model,
\begin{equation}\label{eq:action1}
S=\frac{1}{16\pi}\int d^4 x \sqrt{-g}\left( R +\alpha R^2 \right),\quad \quad \alpha \ge 0.
\end{equation}
The $R^2$ term excites a massive spin-0 mode. Taking the trace of Einstein equation yields the scalar equation  $6\alpha \Box R = R$, which indicates that $R$ behaves as a free massive scalar field. However, it has been proven that $R$ must vanish for black hole solutions\cite{Lu:2015cqa}. 

The Starobinsky gravity is not only the simplest example of $f(R)$ gravity, but also an implicit scalar-tensor theory. Equation \eqref{eq:action1} can be equivalently rewritten in a form that explicitly contains a scalar field,
\begin{equation}
S=\frac{1}{16\pi}\int d^4 x \sqrt{-g}\left( R +\phi R - \frac1 2 \mu ^2 \phi^2  \right),\quad \quad \phi=2\alpha R, \quad \quad \mu^2 = \frac1{2\alpha} > 0.
\end{equation}
Inspired by black hole scalarization theory \cite{Antoniou:2017acq,Silva:2017uqg,Doneva:2017bvd}, Liu \emph{et al}. \cite{Liu:2020yqa} extended the theory with Gauss-Bonnet invariant $E^{\text{GB}}$, confirming that the extended theory allows for black hole solutions with scalar hair, 

\begin{equation}\label{eq:lagrangian}
S=\frac{1}{16\pi}\int d^4 x \sqrt{-g}\left(R + \phi R - \frac{1}{2} \mu^2 \phi^2 + U(\phi)E^{\text{GB}} \right),
\end{equation}
\begin{equation}
E^{\text{GB}}=R^2-4 R_{\mu \nu}R^{\mu \nu}+R^{\mu \nu \alpha \beta} R_{\mu \nu \alpha \beta}.
\end{equation}
The Einstein equation is
\begin{align}\label{eq:totalEoM}
&R_{\mu\nu} - \frac{1}{2} R g_{\mu\nu} 
+ \phi R_{\mu\nu} - \nabla_\mu \nabla_\nu \phi + \Box \phi g_{\mu\nu} 
- \frac{1}{2} \phi R g_{\mu\nu} + \frac{1}{4} \mu^2 \phi^2 g_{\mu\nu} 
- 2R \nabla_\mu \nabla_\nu U \notag\\ 
-& 4 \left( R_{\mu\nu} - \frac{1}{2} R g_{\mu\nu} \right) \Box U 
+ 8 R^\rho{}_{(\mu} \nabla_{\nu)} \nabla_\rho U - 4 R^{\rho\sigma} \nabla_\rho \nabla_\sigma U g_{\mu\nu} 
+ 4 R_\mu{}^{\rho}{}_{\nu}{}^{\sigma} \nabla_\rho \nabla_\sigma U = 0.
\end{align}
The scalar field equation is
\begin{equation}\label{eq:scalarEoM}
R-\mu^2 \phi + U'(\phi)E^{\text{GB}} = 0.
\end{equation}
Combine \eqref{eq:scalarEoM} with the trace of \eqref{eq:totalEoM}, we have
\begin{equation}\label{eq:traceEoM}
3 \Box \phi = \mu^2 \phi - (1+\phi) U'(\phi) E^{\text{GB}} -2R\Box U(\phi) + 4R^{\mu\nu}\nabla_\mu \nabla_\nu U(\phi).
\end{equation}
Clearly, in the case where $U$ vanishes, \eqref{eq:traceEoM} reduces to that of a free scalar field with mass $\frac{\mu}{\sqrt{3}}$, implying that the no-hair theorem applies to black hole solutions in this limit. However, when $U(\phi)\ne 0$, the Gauss-Bonnet term $E^{\text{GB}}$ allows the black hole to support a non-trivial scalar field, thus evading the no-hair theorem.

Since the scalar equation \eqref{eq:scalarEoM} is algebraic, the scalar $\phi$ can be easily expressed as a function of the curvature invariants, $\phi(R, E^{\text{GB}})$, resulting in a purely gravitational action. For the specific linear coupling $U=\beta \phi$, we obtain $\phi = 2\alpha (R +\beta E^{\text{GB}})$, and hence,
\begin{equation}\label{eq:linearU}
S=\frac{1}{16\pi}\int d^4 x \sqrt{-g}\left(R + \alpha (R + \beta E^{\text{GB}})^2  \right).
\end{equation}
In the case of quadratic coupling $U=\frac12 \beta \phi^2$, we have $\phi = 2\alpha R/(1 - 2 \alpha \beta E^{\text{GB}})$, leading to the action 
\begin{equation}\label{eq:squareU}
S=\frac{1}{16\pi}\int d^4 x \sqrt{-g}\left(R + \frac{\alpha R^2}{1-2 \alpha \beta E^{\text{GB}}}  \right).
\end{equation}
%在接下来的内容，我们将如同[xx]一样，围绕第二种理论 \eqref{eq:squareU} 展开计算。我们将基于[xx]的结论做出进一步的突破，并将理论进一步推广到带电的情况。
In the subsequent analysis, we follow the approach of \cite{Liu:2020yqa} and focus on the second model \eqref{eq:squareU}. Building upon this foundation, we extend the previous analysis and further extend the theory to the charged case.
The model  \eqref{eq:squareU}  allows black hole solution with vanishing scalar mode. In that case, the theory reduces to Einstein theory, and we can recover the Schwarzschild solution. Furthermore, we will attempt to construct a new black hole solution containing scalar hairs. In contrast, the first model \eqref{eq:linearU} does not admit the Schwarzschild solution because of the non-vanishing $E^{\text{GB}}$ in \eqref{eq:scalarEoM}.

\section{Numerical Solution Method}
This section is devoted to the numerical construction of static, spherically symmetric, and asymptotically flat black hole solutions. We work with the previously introduced Lagrangian \eqref{eq:lagrangian} , specializing to the case $U=\frac12 \beta \phi^2$. This model represents an extension of the Starobinsky model coupled to the Gauss-Bonnet invariant \eqref{eq:squareU}. We employ the following ansatz:
\begin{align}
ds^2 &= -h(r) dt^2 + \frac{dr^2}{f(r)} + r^2 (d\theta^2 + \sin^2\theta d\varphi^2), \cr
\phi &= \phi(r). \label{ansatz}
\end{align}
While the $\phi=0$ solution of \eqref{eq:scalarEoM} trivially yields the standard Schwarzschild metric ($h=f=1-2M/r$),  our primary interest lies in hairy solutions. For these non-trivial solutions to maintain asymptotic flatness, the behavior of the scalar field is governed by the Yukawa falloff
\begin{equation}\label{eq:Yukawa}
\phi \to \frac{\phi_0}{r} e^{-\frac{\mu  r}{\sqrt{3}}}.
\end{equation}
The presence of the scalar field modifies the metric functions $h(r)$ and $f(r)$. These can be expressed as perturbations around the Schwarzschild solution, with the leading asymptotic behavior taking the form
\begin{align}\label{asymptoticMetric}
h(r)&\to 1-\frac{2 M}{r}- \frac{\phi_0}{r}e^{-\frac{\mu  }{\sqrt{3}}r};\notag\\
f(r)&\to 1-\frac{2 M}{r}+ \phi_0 \left(\frac{\mu }{\sqrt{3}}+\frac{1}{r}\right)e^{-\frac{\mu  }{\sqrt{3}}r}.
\end{align}
The Arnowitt-Deser-Misner (ADM) mass is given by $M$ and is independent of both the scalar hair $\phi_0$ and the theory's couplings $(\beta,\mu)$. While the scalar hair has a negligible effect on the asymptotic structure, it significantly alters the geometry near the horizon. To analyze this, we expand the metric functions and the scalar field near the horizon $r = r_+$:
\begin{align}
h=h_1(r-r_+)+\cdots,\quad f=f_1(r-r_+)+\cdots,\quad \phi=\phi_+ + \phi_1(r-r_+)+\cdots,
\end{align}
\begin{align}\label{eq:LeadingOrder}
f_1 = & \frac{(3 + 3\phi_+ - \beta\mu^2\phi_+^2)r_+}{12\beta\phi_+(\phi_+ + 1)} - \frac{1}{48\beta\phi_+(\phi_+ + 1)} \biggl[16r_+^2(-\beta\mu^2\phi_+^3 + 3\phi_+ + 3)^2  \notag \\
       & - 96\beta\phi_+(\phi_+ + 1)\bigl(4\beta\mu^2\phi_+^3 - 3\mu^2\phi_+^2r_+^2  - 2\phi_+(\mu^2r_+^2 - 6) + 12\bigr)\biggr]^{\frac{1}{2}},\notag \\
\phi_1 =& \frac{4(1 - f_1 r_+)(1 + \phi_+) - \mu^2\phi_+^2 r_+^2}{2f_1(4\beta\phi_+ + r_+^2)}.
\end{align}
To avoid branch cuts, the near-horizon expansion is performed in integer powers. Solving the equations order by order reveals that the solution is characterized by two independent horizon parameters: the radius $r_+$ and the scalar hair $\phi_+ \equiv \phi(r_+)$, which correspond to the asymptotic mass $M$ and scalar charge $\phi_0$. Due to the freedom in time rescaling, the coefficient $h_1$ is, in principle, arbitrary; its value is ultimately fixed by the condition of asymptotic flatness. The temperature and entropy of the black hole are given by
\begin{equation} \label{eq:Wald}
T = \frac{\sqrt{h_1 f_1}}{4\pi}, \quad 
S = \pi r_+^2(1 + \phi_+) + 2\pi\beta\phi_+^2.
\end{equation}
The temperature is computed by requiring the regularity of the Euclidean geometry at the horizon, which fixes the period of Euclidean time. The entropy is derived via the Wald entropy formula \cite{Wald:1993nt}.

Since the equations of motion are difficult to solve analytically, we employ numerical methods to construct black hole solutions with scalar hair. For given coupling parameters $(\mu, \beta)$, we aim to establish the relationship between the asymptotic scalar hair $\phi_0$ and the mass $M$, thereby connecting the horizon data $(r_+, \phi_+)$ to these asymptotic parameters.
The differential equations for $h(r)$, $f(r)$, and $\phi(r)$, derived from the field equations \eqref{eq:totalEoM}-\eqref{eq:scalarEoM} are solved by imposing appropriate boundary conditions.  One standard approach is to use the near-horizon expansions \eqref{eq:LeadingOrder} as initial conditions and integrate outward from the horizon to large $r$. To avoid the singularity at the horizon, the numerical integration is initiated at a point slightly outside $r_+$, using a Taylor expansion to provide the initial data. 
However, this shooting method faces significant challenges. The complexity of the coefficients in \eqref{eq:LeadingOrder} makes it difficult to compute high-order Taylor expansions analytically, limiting the precision of the initial data. More critically, a generic initial value $(r_+,\phi_+)$ could excite non-physical scalar mode $e^{+\mu r/\sqrt3}$. This leads to numerical divergence, meaning that only an extremely fine-tuned set of initial conditions will yield a physical black hole solution that asymptotically approaches the Yukawa falloff \eqref{eq:Yukawa}. This sensitivity poses a major obstacle for numerical analysis, while this method is standard for solutions involving a massive mode and can be effective for systems with simpler coefficients\cite{Lu:2015cqa}.

\begin{figure}[htbp]
  \centering
  \includegraphics[width=0.55\textwidth]{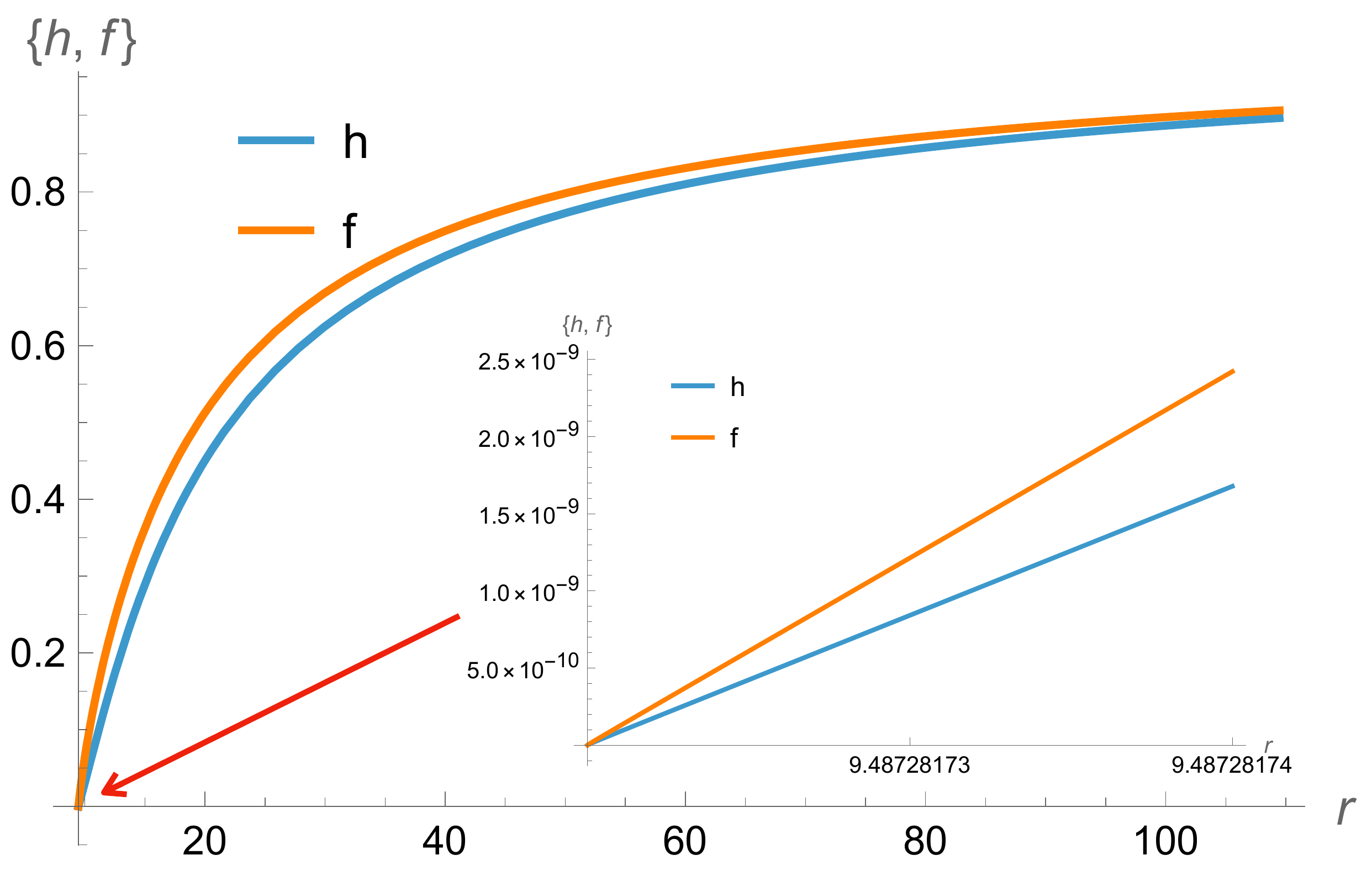}
  \caption{
Metric functions $h(r)$ and $f(r)$ for a scalarized hairy black hole solutions with asymptotic parameters $(M,  \phi_0) = (5.0, 0.3967)$ and theory couplings $(\mu, \beta) = (0.01, 50)$. The event horizon at $r_+ = 9.487$ is identified by the simultaneous vanishing of $h(r)$ and $f(r)$. An obvious distinction from the Schwarzschild black hole, where $h(r)=f(r)$, is that here $h(r)\neq f(r)$ for the hairy case. 
  \label{fig:fh}}
\end{figure}

For our theory, a more effective solution strategy is available. This involves using the asymptotic expansion \eqref{asymptoticMetric} at a sufficiently large radius $r_i$ as a boundary condition and integrating the equations inward. Since the scalar hair decreases exponentially, the high-order terms of expansion can be ignored in numerical calculation. A black hole solution is identified when both metric functions $(h,f)$ vanish at some point, which is then taken as the event horizon $r_+$. In practice, due to finite numerical precision, the functions cannot be made to reach exactly zero. Our numerical results indicate that simultaneously imposing $h(r) < 10^{-6}$ and $f(r) < 10^{-6}$ provides a sufficient criterion for locating the horizon and extracting the black hole's main physical properties. Only when calculating the Hawking temperature, due to the involvement of the slope of the metric function at the event horizon, the initial parameter $(M,\phi_0)$ needs to be adjusted more precisely. Fig. \ref{fig:fh} shows the metric functions $h(r)$ and $f(r)$ for a representative scalarized black hole solution. 

\section{Revisiting and Extending Hairy Black Hole Solutions}
The Schwarzschild metric with $\phi=0$ is always a solution to \eqref{eq:squareU}. Typically, the existence of a new black hole branch implies that the Schwarzschild solution becomes unstable within a specific mass range, leading to the spontaneous growth of scalar hair. To probe this unstable mass interval, we study linear perturbations around a Schwarzschild background of mass $M$. We introduce a test scalar perturbation $\delta \phi \to 0$, assuming its backreaction on the metric is negligible. Substituting $\phi = \delta \phi$ into \eqref{eq:traceEoM} and expanding to linear order in $\delta \phi$, neglecting terms of $\mathcal{O}(\delta \phi^2)$ and higher, we obtain the equation governing the perturbation
\begin{equation}\label{eq:purtuebation}
3 \Box_{(0)} \delta \phi = \mu^2 \delta \phi -  U'(\delta \phi) E^{\text{GB}}_{(0)}.
\end{equation}
where $\Box_{(0)}$ , $ E^{\text{GB}}_{(0)}$ and $R^{\mu\nu}_{(0)}$ are the d’Alembert operator, the Gauss-Bonnet invariant and the Ricci tensor in the Schwarzschild geometry. By fixing the couplings $(\mu, \beta)$, and scanning over the mass $M$, we numerically solve \eqref{eq:purtuebation}. For most masses, only divergent scalar hair modes are excited. 
However, for a specific critical mass $M_0$, a static, asymptotically decaying solution for $\delta \phi$ exists. This zero mode signals a bifurcation point, where a new branch of hairy black hole solutions with nontrivial scalar hair ($\phi \to 0$ at infinity) emerges from the Schwarzschild branch.

\begin{figure}[htbp]
	\centering
	\includegraphics[width=1\textwidth]{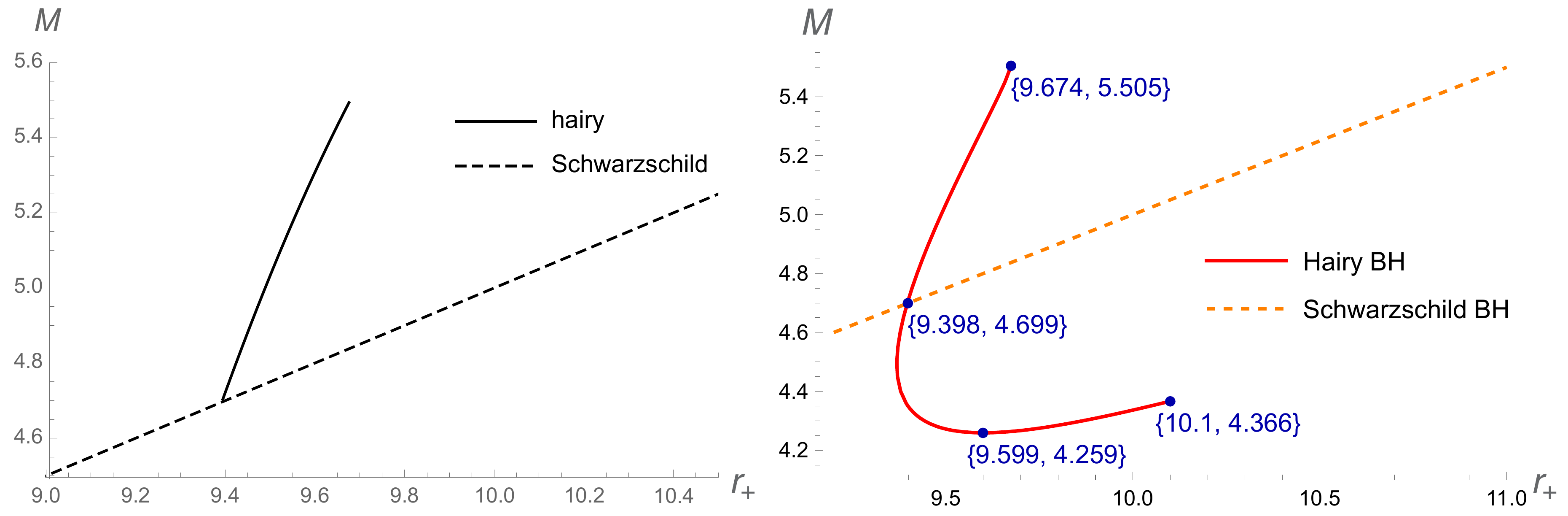}
	\caption{
		The mass as a function of horizon radius is presented for the Schwarzschild black hole and the hairy black hole with $\mu = 0.01$ and $\beta = 50$. In contrast to the known solution branch displayed in the left panel, our new results in the right panel demonstrate that the hairy solution can cross the Schwarzschild black hole solution curve, extending to lower mass regimes. }  \label{fig:2}
\end{figure}

For couplings $\mu=0.01$, $\beta=50$, the linear perturbation analysis yields a critical mass of $M_0 = 4.698$, signaling the potential for hairy black holes. A single, continuous branch of such solutions was first established in \cite{Liu:2020yqa}, as illustrated in the left panel of Fig. \ref{fig:2}. This branch bifurcates from the Schwarzschild solution at $M_0$ and extends up to maximum mass of $M_{\text{max}}=5.505$, with the solutions in this range characterized by positive scalar hair and horizons smaller than Schwarzschild's. Beyond this upper mass limit ($M>M_{\text{max}}$), this particular branch of hairy black holes ceases to exist.

However, the linear perturbation analysis of the scalar field merely indicates that the Schwarzschild black hole develops an unstable mode at $M_0$, and this instability is not restricted to only one sign of the scalar field ($\phi > 0$). Consequently, there is no physical reason to exclude the existence of black hole solutions on the opposite sign ($\phi < 0$). In fact, the perturbative analysis implies that the two sides, $\phi > 0$ and $\phi < 0$, are equivalent at $M_0$. Combined with previous work that has already established the existence of hairy black hole solutions for $\phi > 0$ \cite{Liu:2020yqa}, this equivalence strongly suggests that hairy solutions should also emerge on the $\phi < 0$ side. We have carried out extensive numerical work on this opposite side, and indeed we have successfully found hairy black hole solutions there. Our work completes this picture by characterizing the lower-mass portion of this branch, namely, the region where the solutions possess negative scalar hair and have horizons larger than those of the Schwarzschild black hole, as illustrated in right panel of Fig.~\ref{fig:2}.

\begin{figure}[htbp]
  \centering
  \includegraphics[width=0.6\textwidth]{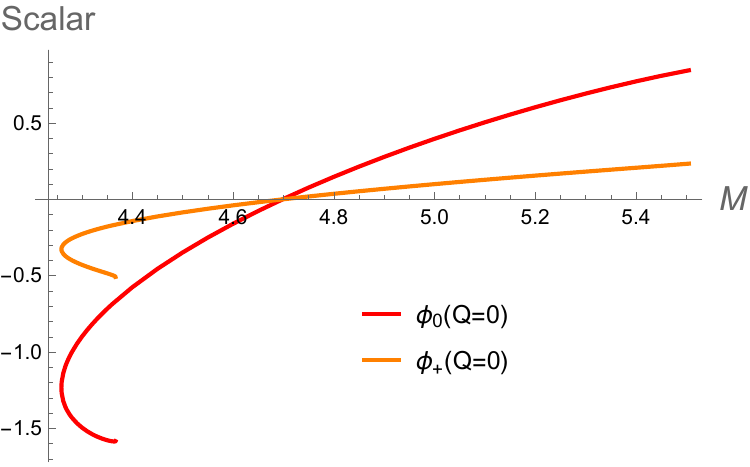}
  \caption{
The figure illustrates that the scalar quantities $(\phi_+, \phi_0)$ are not independent parameters but rather functions of the mass $M$. The existence of multiple distinct scalar parameter values corresponding to a single mass indicates the multibranched structure of hairy black hole solutions.
  \label{fig:3}}
\end{figure}

In the mass--horizon-radius ($M-r_+$) plane, we find that the hairy black hole can cross the hairless Schwarzschild black hole. As the horizon radius $r_+$ decreases, the hairy black hole mass $M$ decreases; however, unlike the Schwarzschild case, $M$ does not tend to zero as $r_+$ shrinks. Instead, the hairy solutions exhibit a minimum horizon radius and a minimum mass, which occur at distinct points. The hairy branch crosses the Schwarzschild one from above, reaches the minimum horizon radius, and then bends rightward to attain a minimum mass. Consequently, the hairy black hole actually consists of two branches, which are smoothly joined at the minimum horizon radius. It should also be noted that, as the hairy branch crosses the Schwarzschild one from above to below, the scalar parameter $\phi_0$ decreases monotonically along the hairy curve, changing from positive to negative and vanishing exactly at the intersection point. And the scalar parameters as a multivalue function of the black hole mass are shown in Fig.~\ref{fig:3}.

%As shown in the right panel of Fig. \ref{fig:2}, hairy black hole solutions exist for masses $M > 4.259$. For a given mass within a certain range, we find two distinct hairy solutions characterized by different horizon radii and scalar hair values $\phi_0$. 
%The branch with the larger radius corresponds to a more negative value of $\phi_0$, indicating a strong correlation between the horizon size and the scalar hair, as further illustrated in
%The branch solution with a smaller radius exists for $4.259 < M < 5.505$. At lower masses, its radius is larger than that of the Schwarzschild black hole; however, as the mass increases beyond $M_0$, its radius will be smaller than the Schwarzschild radius, thus evolving into the solutions discovered in previous studies.
%The branch with a larger radius ceases to exist for $M > 4.366$. 

\begin{figure}[htbp]
  \centering
  \includegraphics[width=1\textwidth]{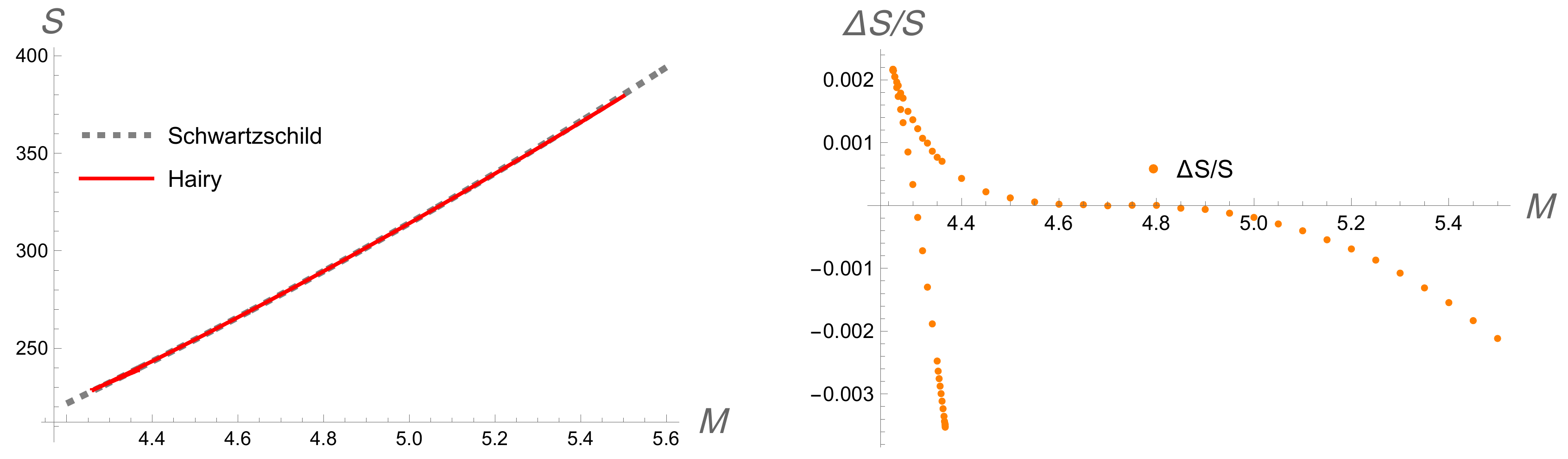}
  \caption{
The left panel compares the entropy/mass relation for the hairy black hole with ($\mu = 1/100$, $\beta = 50$) and the Schwarzschild black hole. The right panel plots the entropy difference $\Delta S = S_{\text{hairy}} - S_{\text{S}}$, showing a relative deviation of less than $0.36\%$.
  \label{fig:entropy uniformity}}
\end{figure}

Furthermore, the intriguing phenomenon that, for a fixed mass $M$, the entropy of a hairy black hole is nearly identical to that of the Schwarzschild black hole persists for the new branches we have constructed, as shown in Fig.~\ref{fig:entropy uniformity}. In the left panel, the entropy--mass curves for the Schwarzschild black hole and the two hairy branches are almost indistinguishable, overlapping nearly completely. The right panel displays the relative difference between the entropy of the hairy black hole and that of the Schwarzschild counterpart as a function of the black hole mass. From this panel, it is evident that the entropy deviations for both hairy branches are exceedingly small at the same mass.

Inspection of the entropy formula \eqref{eq:Wald} shows that the entropy of hairy black hole receives contributions from both the horizon area and the scalar field. The near equality of entropy between the hairy and Schwarzschild black holes therefore implies a nontrivial compensation between these two contributions.

Not only is the entropy consistent, but the first law of thermodynamics for the hairy black hole also takes the same form as in Einstein's theory, namely, $\text{d}M = T\text{d}S$. 
We note that the massive scalar charge $\phi_0$ does not contribute to the first law of black hole thermodynamics, as its thermodynamic conjugate variable corresponds to a divergent solution component and must consequently be constrained to vanish.

\section{Charged Hairy Black Hole Solutions} 
In this part, we will further extend the theory \eqref{eq:lagrangian} to explore electrically charged hairy black holes by introducing the Maxwell field  into the action, 
\begin{equation}\label{eq:ChargedLagrangian}
S=\frac{1}{16\pi}\int d^4 x \sqrt{-g}\left(R + \phi R - \frac{1}{2} \mu^2 \phi^2 + U(\phi)E^{\text{GB}} -  F_{\mu \nu}F^{\mu \nu} \right),
\end{equation}
where $F_{\mu\nu}=\partial_{[\mu}A_{\nu]}$ is the Maxwell tensor. 
The Einstein equation then transforms into 
\begin{align}\label{eq:ChargedEoM}
&R_{\mu\nu} - \frac{1}{2} R g_{\mu\nu} 
+ \phi R_{\mu\nu} - \nabla_\mu \nabla_\nu \phi + \Box \phi g_{\mu\nu} 
- \frac{1}{2} \phi R g_{\mu\nu} + \frac{1}{4} \mu^2 \phi^2 g_{\mu\nu} \notag\\
-&2F_{\mu\alpha}F_\nu{}^{\alpha} + \frac12g_{\mu\nu}F_{\alpha\beta}F^{\alpha\beta}
-2R \nabla_\mu \nabla_\nu U 
- 4 \left( R_{\mu\nu} - \frac{1}{2} R g_{\mu\nu} \right) \Box U\notag \\
+& 8 R^\rho{}_{(\mu} \nabla_{\nu)} \nabla_\rho U - 4 R^{\rho\sigma} \nabla_\rho \nabla_\sigma U g_{\mu\nu} 
+ 4 R_\mu{}^{\rho}{}_{\nu}{}^{\sigma} \nabla_\rho \nabla_\sigma U = 0.
\end{align}
The Maxwell equation is given by
\begin{equation}\label{eq:MaxwellEoM}
\partial_\mu(\sqrt{-g}F^{\mu\nu})=0.
\end{equation}
%电荷的引入对标量场方程没有影响，eq. \eqref{eq:scalarEoM} 没有改变，同样的，标量场的渐进行为\eqref{eq:Yukawa} 也没有改变。又因为 Maxwell tensor 是无迹的，因此Einstein 方程的迹不会改变，从而使得\eqref{eq:traceEoM} 同样保持原本的形式。另外，电荷也不会改变温度和熵的公式。
The introduction of electric charge leaves the scalar field equation \eqref{eq:scalarEoM} unaltered. Consequently, the asymptotic behavior of the scalar field, given by \eqref{eq:Yukawa}, also remains identical. Furthermore, since the Maxwell tensor is traceless, it does not contribute to the trace of the Einstein equations. This ensures that the traced equation \eqref{eq:traceEoM} retains the same form. Here, we still adopt the static and spherical metric ansatz (\ref{ansatz}), with electric potential\[
 \quad A_\nu = (-A_0(r),0,0,0).
\]
Maxwell's equation implies that the electric potential can be expressed via the metric functions
\[
\quad A_0'(r)=\sqrt{\frac{h(r)}{f(r)}}\frac{Q}{r^2} \,,
\]
here, $Q$ is the integration constant related with charge. It is worth noting that the expressions for the Hawking temperature and entropy have the same form as in the neutral case when written in terms of the metric functions and scalar field.
%在带电情形下，度规函数的无穷远渐进行为应写为RN时空的微扰。由于RN时空不是真空解，text scalar field 的微扰方程中含$R^{\mu\nu}_{(0)}$的项no vanish，微扰方程变为
For the charged case, the asymptotic behavior of the metric functions corresponds to that of perturbations about the Reissner--Nordström solution,
\begin{align}\label{ChargedasymptoticMetric}
h(r)&\to 1-\frac{2 M}{r}+\frac{Q^2}{r^2}+ \phi_0\left(-\frac{3 M}{r^2}+\frac{2 Q^2}{r^3}+\frac{1}{r}\right)e^{-\frac{\mu  r}{\sqrt{3}}};\notag\\
f(r)&\to 1-\frac{2 M}{r}+\frac{Q^2}{r^2}+ \phi_0 \left(\frac{\mu }{\sqrt{3}}-\frac{M}{r^2}-\frac{2 \mu  M}{\sqrt{3} r}+\frac{\mu  Q^2}{\sqrt{3} r^2}+\frac{1}{r}\right)e^{-\frac{\mu  r}{\sqrt{3}}}.
\end{align}
%由于RN时空不是真空解，text scalar field 的微扰方程中含$R^{\mu\nu}_{(0)}$的项no vanish，微扰方程变为
Since the Reissner-Nordström spacetime has a non-vanishing Ricci tensor, the perturbation equation of test scalar field becomes
\begin{equation}\label{eq:Chargedpurtuebation}
3 \Box_{(0)} \delta \phi = \mu^2 \delta \phi -  U'(\delta \phi) E^{\text{GB}}_{(0)}+ 4R^{\mu\nu}_{(0)}\nabla_\mu \nabla_\nu U(\delta \phi).
\end{equation}
where $\Box_{(0)}$ , $ E^{\text{GB}}_{(0)}$ and $R^{\mu\nu}_{(0)}$ are the D’alambert operator, the Gauss-Bonnet invariant and the Ricci tensor in the Reissner-Nordström geometry.

\begin{figure}[htbp]
  \centering
  \includegraphics[width=0.7\textwidth]{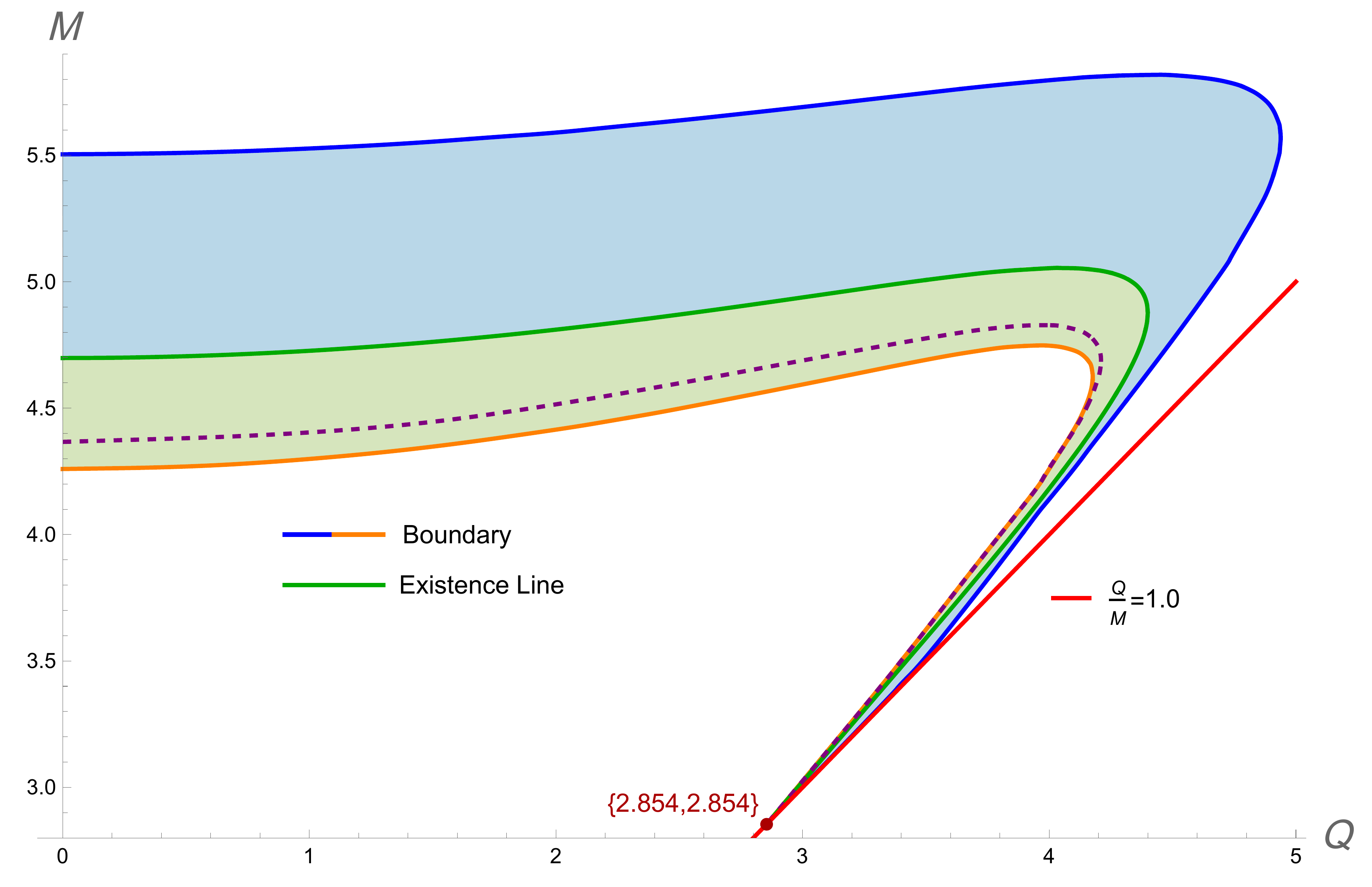}
  \caption{
This figure shows the existence domain in the $\{Q,M\}$ parameter space where hairy black hole solutions exist for the theory with coupling parameters $\mu = 0.01$, $\beta = 50$. The green existence line denotes the bifurcation points of two kinds of black hole solutions, extending to $(2.85,2.85)$. All solutions satisfy $Q/M < 1$.}   \label{fig:area}
\end{figure}

%引入电荷后，我们不仅需要考察带毛黑洞存在的质量范围，还要考虑电荷区间。我们发现，在如图5所示的$\{Q,M\}$参数空间中，带毛黑洞解的分支点随着电荷的变化形成了一条曲线，在图中我们用绿色曲线标记。带毛黑洞存在的参数区域分布在绿色曲线附近，分别用淡蓝色和淡绿色标记。因此我们将绿色曲线称为存在线，标记了颜色的区域称为存在域。带毛黑洞在存在线上会退化为RN黑洞，此时标量毛vanish。淡蓝色存在域上的带毛黑洞的标量荷$\phi_0>0$，视界半径小于RN黑洞半径，淡绿色存在域与之相反。存在域的边界用蓝色和橙色曲线标记，分别称为外边界和内边界。在存在域的一些特定区间内，存在两个不同的带毛黑洞解，类似图二右图下端的情况，这样的区域分布于虚线和内边界之间。存在域最下端的部分黑洞解近似于极端黑洞。
With the introduction of electric charge, we must consider not only the mass range but also the charge interval in which hairy black holes exist. We find that in the $(Q,M)$ parameter space, as illustrated in Fig. \ref{fig:area}, the bifurcation points of the hairy black hole solutions form a continuous curve with varying charge, marked by the green curve in the plot. The region in parameter space where hairy black holes exist is distributed around this green curve and is shaded in light blue and light green. Accordingly, we refer to the green curve as the \textit{existence line} and the colored region as the \textit{existence domain}.
On the existence line, the hairy black holes reduce to Reissner-Nordström black holes, and the scalar hair vanishes. Within the light blue domain, the hairy black holes possess a positive scalar charge $\phi_0 > 0$ and have horizon radii smaller than those of the comparable Reissner-Nordström black hole. In contrast, the solutions in the light green domain exhibit the opposite behavior. The boundaries of the existence domain are marked by blue and orange curves, termed the \textit{outer} and \textit{inner} boundaries, respectively. Solutions at the very bottom of the domain approach the extremal limit. For all solutions, including the Reissner--Nordstr\"om black hole, the charge-to-mass ratio is bounded by $Q/M \le 1$.

Here we want to highlight a novel feature of the scalarized black hole solutions within our framework: the hairy black hole geometries split into multi-distinct branches. This multi-branch structure has never been observed in prior studies of scalarized spherically symmetric black holes in other modified gravity models. One terminus of some branches lies inside the  existence domain. We mark these critical endpoints and connect them via the purple dashed curves plotted in Fig. \ref{fig:area}. 

%In certain specific intervals within the existence domain distributed between the dashed line and the inner boundary, two distinct hairy black hole solutions coexist for the same $(Q, M)$ parameters, analogous to the lower branch structure shown in the right panel of Fig. \ref{fig:2}. 

\begin{figure}[!ht]
  \centering
  \includegraphics[width=1\textwidth]{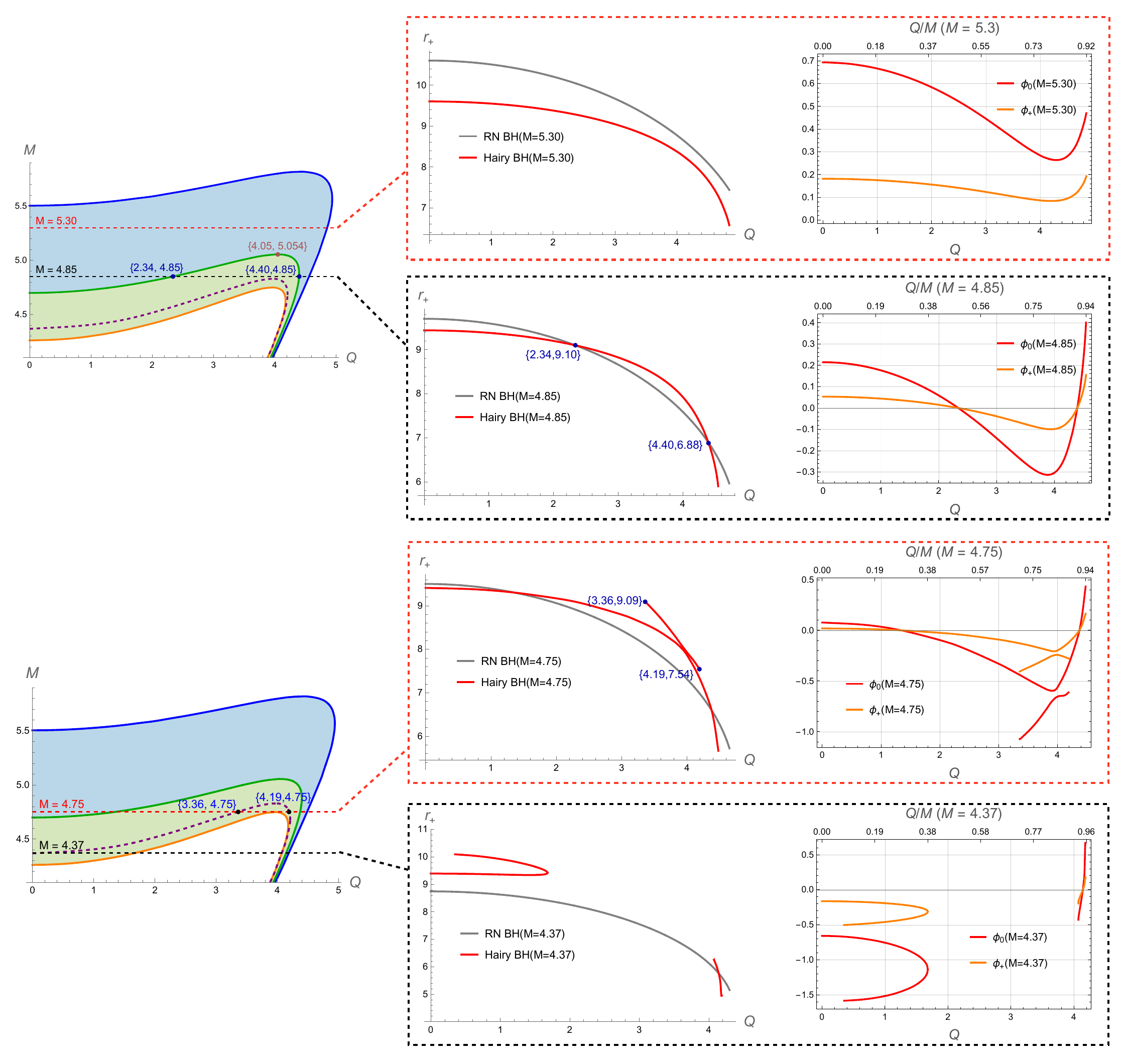}
  \caption{
%该图选取了两个特定质量以研究固定质量情况下黑洞解随电荷的变化关系，。引入电荷后，标量毛与黑洞的半径差仍然同步变化。
This figure investigates the charge dependence of hairy black hole solutions evaluated at four distinct fixed masses. The left panels display constant-mass contours alongside the solution existence domain, with critical intersection points highlighted.
The right panels present the corresponding charge dependence of horizon radii and scalar hair quantities. The endpoints of hairy solution curves terminate either on the domain boundary curves or the purple dashed line. When a solution curve intersects the purple dashed line, the associated hairy black hole solutions exhibit a structure comprising two disconnected branches; otherwise, only a single continuous smooth solution curve emerges.
}
  \label{fig:fixM} 
\end{figure}

By fixing one of the parameters $(Q,M)$ and varying the other, we can reveal intriguing properties of the hairy solutions. When the mass $M$ is fixed, the scalar hair does not vary monotonically with the charge. In Fig. \ref{fig:fixM}, we investigate this by studying the charge dependence of the black hole solutions for several specific mass values. The intersection points between the constant-mass line and the curves of the existence domain are key to understanding the solution structure.

For large masses, constant-mass contours no longer intersect the purple dashed line, which implies a unique hairy black hole solution. When the mass exceeds a critical threshold, the constant-mass curve fails to intersect the existence boundary, and the scalar hair profile possesses no zero crossing. Taking $M=5.3$ as an illustrative example, the corresponding constant-mass curve intersects neither the existence boundary nor the purple dashed line; accordingly, only a single solution branch survives, and the scalar hair profile contains no zero crossing. For $M=4.85$, the constant-mass line intersects the existence line at two distinct points, which yields a scalar field profile that crosses zero twice. This implies that the hairy black hole solution reduces to the Reissner–Nordström solution at these two characteristic charge values.

For small masses, constant-mass contours intersect the purple dashed line, giving rise to multiple solution branches. For $M=4.75$, the constant-mass contour extends from the left, crosses the purple dashed line, and terminates at the boundary line. Two solution branches arise in this case, whose charge intervals partially overlap, as visualized in Figure 6. For $M=4.37$, the constant-mass contour splits into two disconnected segments. The first segment originates on the  left, intersects the purple dashed line once, and terminates at the inner existence boundary. The second segment starts from the purple dashed line and ends on the outer boundary line. This geometric structure yields two entirely distinct families of hairy black hole solutions.

\begin{figure}[!ht]
  \centering
  \includegraphics[width=1\textwidth]{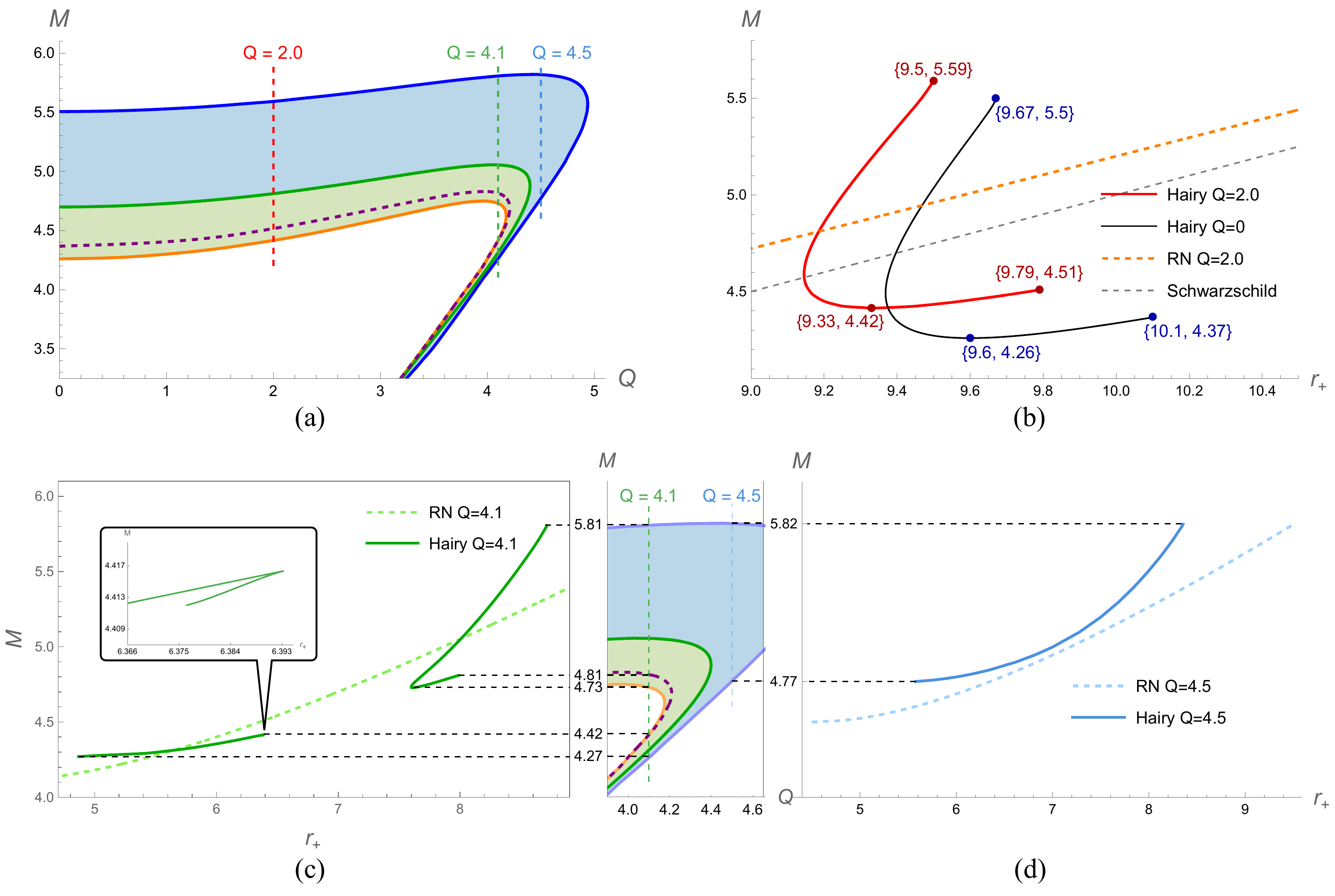}
  \caption{
  %该图选取了五个特定的电荷值以研究固定电荷情况下黑洞解随质量的变化关系，左图中等电荷线与存在域存在交集，使得右侧实线描绘的带毛解出现。在电荷改变的过程中，带毛解的质量-视界半径关系也发生显著的改变。  
  This figure presents hairy black hole solutions for three distinct fixed charge values and illustrates their mass dependence at constant charge. In panel (a), the segments of constant-charge contours residing within the solution existence domain correspond to the hairy black hole solutions represented by solid curves in panels (b)–(d). The mass-horizon radius relation of these hairy solutions undergoes substantial transitions as the charge parameter varies.
    }
  \label{fig:fixQ1}
\end{figure}

When fixing the charge $Q$ and scanning the admissible mass range for hairy black holes, we classify the resulting configurations into three distinct categories. The first case, illustrated in Fig. \ref{fig:fixQ1}(a), corresponds to a comparatively small charge value $(Q=2.0)$. The curve describing hairy black hole solutions in Fig. \ref{fig:fixQ1}(b) bears strong resemblance to the uncharged counterpart. The upper and lower mass limits of these solutions are exactly set by the intersections between the constant-charge contour and the outer and inner boundaries of the existence domain, respectively.

%当我们固定Q并探究黑洞存在的质量范围，结果可以总结为三类。第一种情况描绘与图7(a)，当电荷较小$(Q=2.0)$时，带毛黑洞解的曲线与不带电情况下的曲线很相似，解的质量上下界正是等电荷线与存在域外边界和内边界的坐标。当电荷继续增大，使得等电荷线右移至与存在域内外边界各有两个交点时，便是由图7(b)所描绘的第二种情况。此时等电荷线有两段区间处于存在域内，在图像中体现为质量不连续的两部分解。两部分解都呈现勾形，勾的短臂的质量范围坐落于存在域的虚线与外边界之间。可以设想，当等电荷线在存在域中继续右移，直到与内边界没有交点，第二种情况中分离的两部分解便会连接在一起，这就是第三种情况。我们将三种情况并列展示于图7(d)中。

As the charge grows larger, the constant-charge contour shifts rightward and eventually intersects the inner and outer existence boundaries at two separate points apiece. This constitutes the second solution class, as displayed in Fig. \ref{fig:fixQ1}(c). Here, the constant-charge contour traverses two disconnected intervals inside the existence domain, which correspond to two distinct separate branches of hairy black hole solutions in the parameter diagram. Each branch adopts a hooked geometry. The upper branch originates from the upper outer boundary, crosses inward through the existence line, reaches the inner boundary, then curves back and terminates on the purple dashed line. Meanwhile, the shorter branch follows the reverse trajectory: it starts at the lower outer boundary, crosses upward across the existence line, intersects the inner boundary, bends back, and finally ends on the purple dashed line.

\begin{figure}[!ht]
  \centering
  \includegraphics[width=1\textwidth]{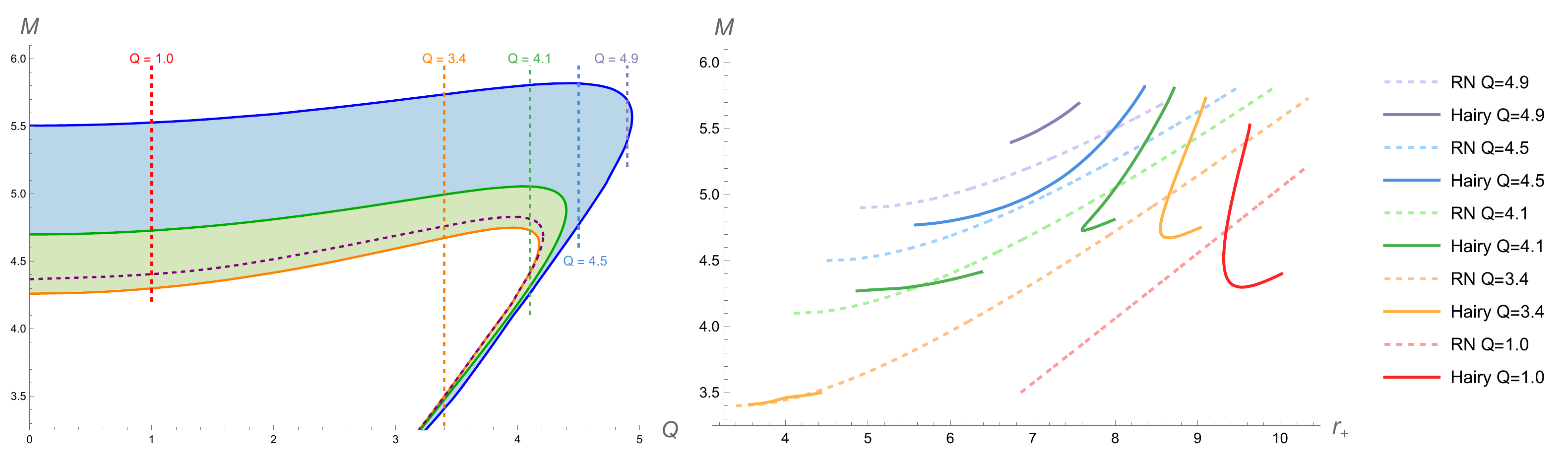}
  \caption{
  %图中选取了5个不同的电荷值来展示三种黑洞解配置的变化，当电荷较小时只存在一支($Q=1.0$)，电荷稍大后第二支解出现，两支解相距较远($Q=3.4$)，并随着电荷增大逐渐链接为一支解($Q=4.1-4.5$)，最终解的存在范围随着存在域收窄。  
  Summary of hairy black hole solutions in the mass-horizon radius plane for five selected charge values, demonstrating the evolution across three distinct solution configurations. A single branch is present at small charge ($Q = 1.0$); a second separate branch emerges at $Q = 3.4$, clearly detached from the original one. As the charge rises from $Q = 4.1$ to $4.5$, the two branches gradually converge and ultimately coalesce into a single continuous solution branch. For further increased charge magnitudes, the admissible mass range of solutions contracts alongside the shrinking existence domain.
  }
  \label{fig:summary}
\end{figure}

As the charge further increases, the constant-charge contour no longer intersects the purple dashed line, leaving only a single solution branch. When the charge becomes sufficiently large, the constant-charge contour fails to cross the existence line, too. Under such circumstances, the scalar profile of this unique branch contains no zero crossing, as illustrated in Fig. \ref{fig:fixQ1}(d). We summarize the configurations corresponding to all three categories of solutions in a single composite plot shown in Fig. \ref{fig:summary}. 

From Fig. \ref{fig:summary}, one can observe that as the charge increases from small to large values, the hairy black hole solution configuration evolves from a single hooked branch to two disconnected hooked branches. Upon further growth of the charge, these two separate hooked segments connect with each other. Nevertheless, the transition process from the double-branch hooked configuration to the single-branch solution is not clearly manifested in this figure. We will conduct a more detailed investigation of the transition regime between these two geometric configurations in subsequent analysis.

\begin{figure}[!ht]
	\centering
	\includegraphics[width=1 \textwidth]{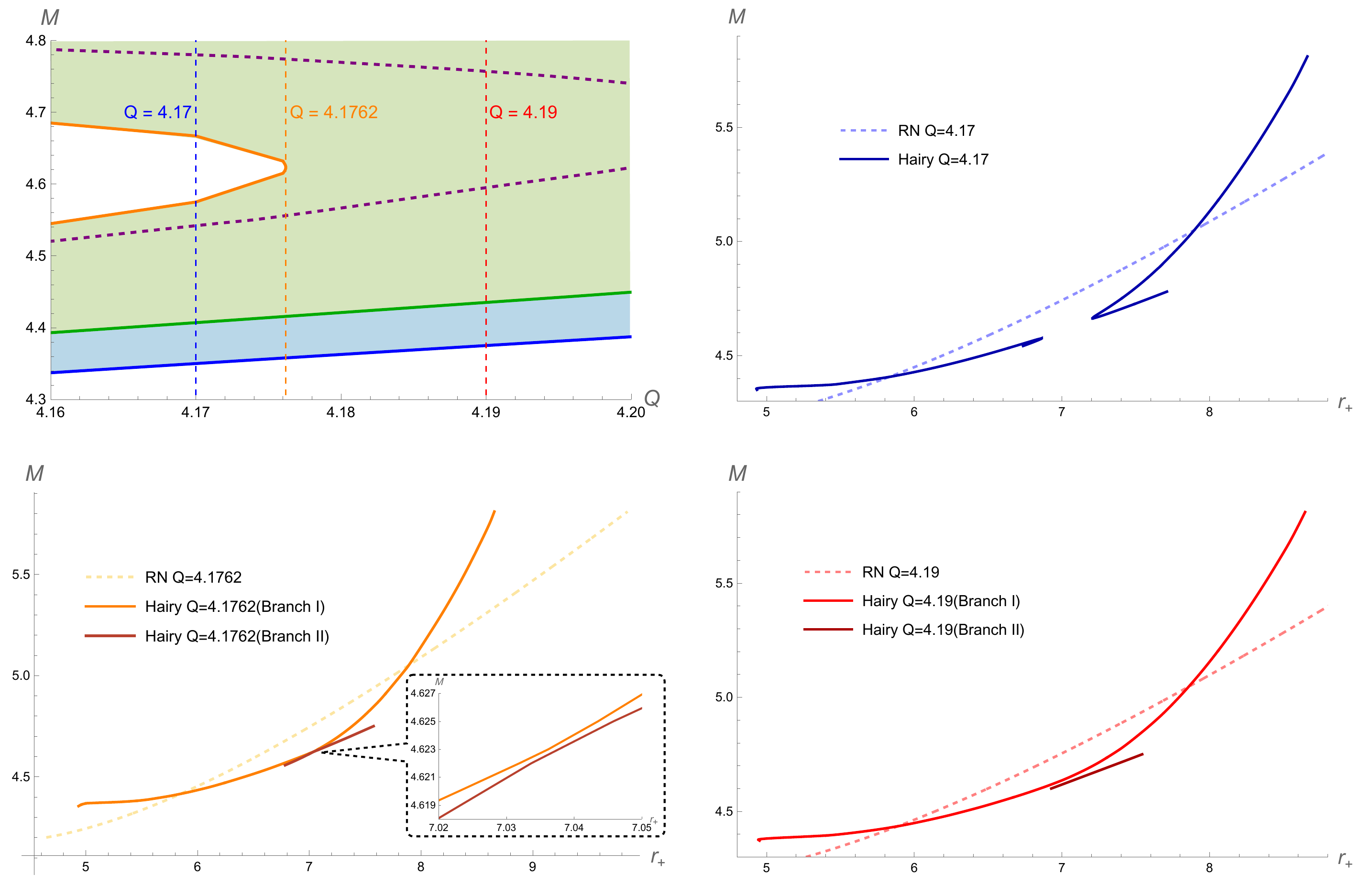}
	\caption{
		%该图显示了disjoint 的黑洞解逐渐连接的过程，两部分钩型的解连接之后，自然地演化成了两个分立的分支。
		This figure depicts the coalescence process of two disconnected hairy black hole solution branches. As the charge parameter increases, the long segments of the two hook-shaped branches join together, as do their short segments, forming one long branch and one short branch. Upon further increasing the charge, the short branch progressively shrinks in extent.
		}
	\label{fig:fixQ2}
\end{figure}

As illustrated in Fig. \ref{fig:fixQ2}, two hook-shaped solution branches are clearly observed at $Q=4.17$, with their hooked segments considerably shortened. When the charge grows to $Q=4.7162$, the long arms of the two original hooked branches merge with one another, as do their short arms, which results in one extended branch and one truncated branch. At this charge value, the constant-charge contour no longer intersects the inner boundary line and instead intersects the purple dashed line at two separate points. Two distinct solution branches remain present, though all hooked geometric features vanish entirely. The longer branch both originates and terminates on the outer  boundary line, while the shorter branch lies entirely between two points on the purple dashed line, starting and ending on this curve. As the charge continues to increase, the shorter branch progressively contracts before disappearing entirely, such that only the long solution branch survives.

%事实上，图7(d)中代表第三种情况的蓝色曲线，看起来更像是由钩形曲线的长臂连接而成的，图7(d)并没有体现短臂的演化。我们将第二种情况和第三种情况之间的过渡更细致地展示于图8.

%当分离的两部分解逐渐将要连接时$(Q=4.17)$，两段解的钩形越发锋利。使等电荷线右移至与内边界没有交点$(Q=4.1762)$，两部分解完成连接，原本钩形的长臂和短臂各自连接形成了一长一短两个分支。事实上，短臂分支的质量上下界正是等电荷线与存在域中虚线的交点坐标。继续增大电荷，短臂分支快速变短最终消失。

We now proceed to analyze the thermodynamics of these hairy black hole solutions. We observe that, for fixed black hole mass and charge, the entropy of the hairy black hole nearly coincides with that of the Reissner–Nordström (RN) black hole. This behavior mirrors the neutral case discussed in the preceding section.

\begin{figure}[!ht]
  \centering
  \includegraphics[width=1\textwidth]{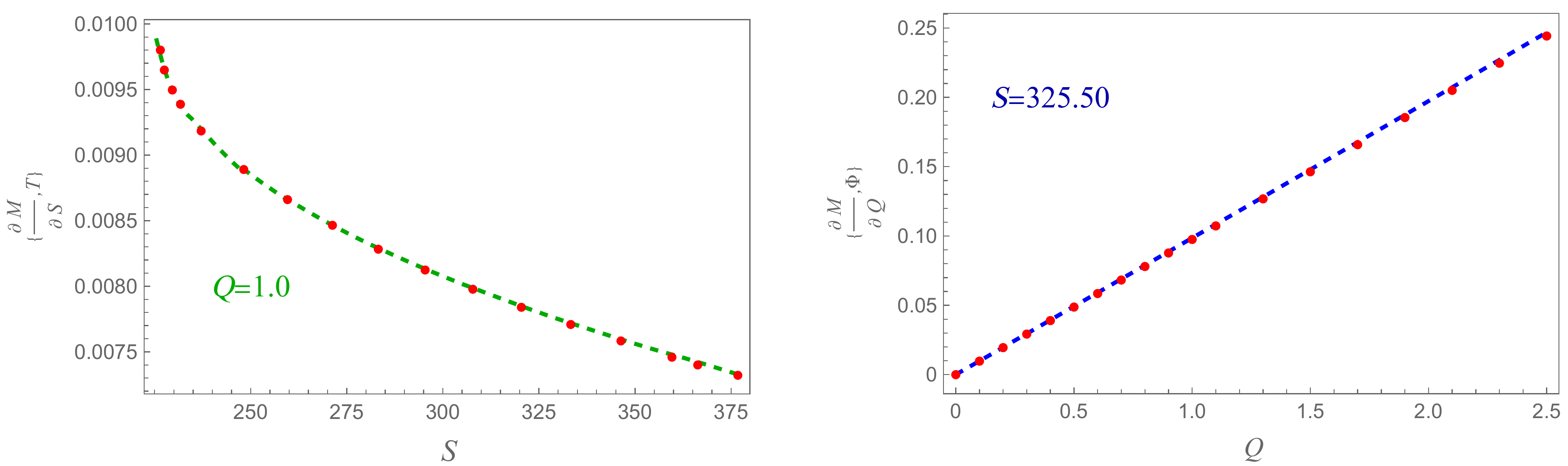}
  \caption{
%左图验证了第一个Maxwell关系，绿色曲线代表$\left({\partial M}/{\partial S} \right)_Q $，红点标出了温度$T$的值。右图验证了第二个Maxwell关系，蓝色曲线代表$\left( {\partial M}/{\partial Q} \right)_S$，红点标出了电势$\Phi$的值。
The left panel verifies the first Maxwell relation. The green dashed line represents $(\partial M/\partial S)_Q$ derived from the fitted $M(S)$ function, while the red dots mark the independently computed temperature $T$. The right panel verifies the second Maxwell relation. The blue curve shows $(\partial M/\partial Q)_S$ from the fitted $M(Q)$ function, and the red dots denote the electrostatic potential $\Phi$.
 }
  \label{fig:1stlaw}
\end{figure}

%在引入电荷后，带毛黑洞第一定律的形式保持不变
The first law of thermodynamics maintains its standard form  after introducing electric charge,
\begin{equation}\label{eq:1stLaw}
\mathrm{d} M = T \mathrm{d} S + \Phi \mathrm{d} Q.
\end{equation}
%我们通过验证如下两个Maxwell关系验证了这一点
 where $\Phi$ is the electrostatic potential at the horizon. We have numerically verified the consistency of this law by checking the corresponding Maxwell relations,
\begin{equation}\label{eq:MaxwellRela}
\left( \frac{\partial M}{\partial S} \right)_Q = T, \quad \left( \frac{\partial M}{\partial Q} \right)_S = \Phi .
\end{equation}
%我们固定电荷的值为$Q=1$，将质量拟合为熵的函数，进而做出$\left({\partial M}/{\partial S} \right)_Q $的曲线。再将$(S,T)$的点标在图上，对比$T$与$\left( {\partial M}/{\partial S} \right)_Q $的值，如果能够匹配便证明了第一个Maxwell关系。同理，在固定熵$S=325.5$的情况下，将质量拟合为电荷的函数$M(Q)$，便可对比$\left( {\partial M}/{\partial Q} \right)_S$和$\Phi$的值。我们将相应的图像展示于Fig. \ref{fig:1stlaw}.
The verification proceeds as follows. First, with the charge fixed at $Q=1.0$, we obtain a fitted function $M(S)$ from our numerical solutions. Its derivative $(\partial M/\partial S)_Q$ is then compared with the independently computed temperature $T$ at the same $S$ values. The close agreement between these two quantities, as shown in the left panel of Fig. \ref{fig:1stlaw}, confirms the first relation. Second, to verify the second relation, we fix the entropy at $S = 325.5$ and fit the mass as a function of charge, $M(Q)$. The derivative $(\partial M/\partial Q)_S$ is then compared with the corresponding potential $\Phi$, with the result displayed in the right panel of Fig. \ref{fig:1stlaw}. The excellent match in both cases provides strong numerical support for the first law \eqref{eq:1stLaw}.

 \newpage

\section{Conclusions}

In this paper, we revisit the spontaneous scalarization of black holes within Starobinsky gravity augmented with  Gauss–Bonnet combination. This theory admits the Schwarzschild black hole as a trivial vacuum solution. Prior literature has demonstrated that scalarized hairy black holes can bifurcate off the Schwarzschild background, with all such hairy solutions residing solely on one side of the Schwarzschild existence curve. Through a more comprehensive analysis, we discover that hairy black hole geometries can cross the Schwarzschild black hole and populate parameter space on both sides of the Schwarzschild solution. Additionally, we uncover an unusual characteristic: the scalarized black hole solutions split into two separate branches sharing a common minimal event horizon radius, at which these two branches connect smoothly. To the best of our knowledge, such a smoothly connected multi-branch structure has never been reported in existing studies of static, spherically symmetric black hole scalarization.

We further generalize this extended Starobinsky gravity by incorporating a Maxwell field to investigate charged hairy black hole configurations. The Reissner–Nordström black hole is naturally a solution of the theory, and we perform a linear perturbation analysis on the RN background to locate the bifurcation point at which charged scalarized black holes first emerge. In relevant existing studies, the scalar charge is not a free independent parameter, and the associated scalar hair is referred to as secondary hair. Accordingly, each charged hairy black hole solution is characterized by two conserved charges: the ADM mass $M$ and electric charge $Q$. We map out the full existence domain of charged hairy black holes within the two-dimensional $(M,Q)$ parameter space.

In the neutral limit, we previously uncovered a multi-branch geometric structure for hairy black hole solutions. This multi-branch feature persists when electric charge is introduced; moreover, charged configurations exhibit an entirely new pattern featuring an extra disconnected solution branch. For these two disconnected families of solutions, their terminal endpoints either lie on the boundaries of the existence domain or reside in the interior region of the parameter space. These interior endpoints collectively form a continuous curve, which we plot as the purple dashed line in Fig. \ref{fig:area}. Any constant-mass or constant-charge contour that intersects this purple dashed line yields two separate disconnected solution branches. In contrast, contours without such intersections correspond to a single smooth solution branch only.

We conduct an extremely thorough investigation into the transition mechanism between single smooth solutions and pairs of disconnected branches, revealing rich and intricate geometric substructures underlying this transformation. 
 The multi‑branch structure suggests that the thermodynamic space of black holes contains a wealth of phase diagrams. Furthermore, the transition between single and disconnected branch solutions may correspond to a specific type of black hole phase transition, which we intend to explore further in future studies to achieve new insights in this direction.

\section{Acknowledgement}
 This work is supported in part by NSFC (National Natural Science Foundation of China) Grant No. 12575061, Tianjin University Graduate Liberal Arts and Sciences Innovation Award Program (2023) No. B1 2023-005 and Tianjin University Self-Innovation Fund Extreme Basic Research Project Grant No. 2025XJ21-0007.

\end{document}